\documentclass[
aps,
nofootinbib,
superscriptaddress,
tightenlines,
notitlepage,
twocolumn,
showpacs,
floatfix,
]{revtex4-1}
\usepackage{amssymb,amsmath,bm,tensor,braket}
\usepackage[colorlinks]{hyperref}

\usepackage{graphicx}% Include figure files
\usepackage{dcolumn}% Align table columns on decimal point
\usepackage{bm}% bold math
\usepackage{xcolor}
\newcommand{\PlanckMass}{M_{\rm Pl}}

\newcommand{\Tabriz}{\affiliation{Faculty of Physics, University of Tabriz,
Tabriz 51666-16471, Iran}}
\newcommand{\KIAA}{\affiliation{Kavli Institute for Astronomy and
Astrophysics, Peking University, Beijing 100871, China}}
\newcommand{\NAOC}{\affiliation{National Astronomical Observatories,
Chinese Academy of Sciences, Beijing 100012, China}}

\begin{document}

\preprint{APS/123-QED}

%---------------------------------------------------------------------
\title{Cosmological perturbations in Gauss-Bonnet quasi-dilaton massive gravity}
% Force line breaks with \\ %\thanks{A footnote to the article title}%

\author{Amin Rezaei Akbarieh}\email{am.rezaei@tabrizu.ac.ir}\Tabriz
\author{Sobhan Kazempour}\email{s.kazempour@tabrizu.ac.ir}\Tabriz
\author{Lijing Shao}\email{lshao@pku.edu.cn}\KIAA\NAOC
% \affiliation{${}^1$
% Faculty of Physics, University of Tabriz, Tabriz, Iran}%
% %\collaboration{MUSO Collaboration}%\noaffiliation
%  %\homepage{http://www.Second.institution.edu/~Charlie.Author}
% \affiliation{${}^2$Kavli Institute for Astronomy and
% Astrophysics, Peking University, Beijing
% 100871, China}%
% \affiliation{${}^3$National Astronomical Observatories,
% Chinese Academy of Sciences, Beijing
% 100012, China}
% %\author{Delta Author}
% %\affiliation{%
%  %Authors' institution and/or address\\
%  %This line break forced with \textbackslash\textbackslash
% %}%


\date{\today}% It is always \today, today,

%---------------------------------------------------------------------
\begin{abstract}
We present the cosmological analysis of the Gauss-Bonnet quasi-dilaton massive gravity theory. This offers a gravitational theory with a nonzero graviton mass. We calculate the complete set of background equations of motion. Also, we obtain the self-accelerating background solutions and we present the constraints on parameters to indicate the correct sign of parameters. In addition, we analyse tensor perturbations and calculate the mass of graviton and find the dispersion relation of gravitational waves for two cases. Finally, we investigate the propagation of gravitational perturbation in the Friedman-Lema\^itre-Robertson-Walker cosmology in the Gauss-Bonnet quasi-dilaton massive gravity.
%\begin{description}
%\end{description}
\end{abstract}


\maketitle

%\tableofcontents

%---------------------------------------------------------------------
\section{\label{sec:intro}Introduction}
%---------------------------------------------------------------------

It is clear that the general theory of relativity has great successes in
Solar System tests \cite{Will:2014kxa, Reynaud:2008yd, Everitt:2011hp} and
various astronomical observations \cite{Berti:2015itd, Wex:2014nva,
Shao:2016ezh}. However, there remain open questions in gravity, cosmology
and particle physics, such as the hierarchy problem
\cite{ArkaniHamed:1998rs}, the cosmological constant problem
\cite{Weinberg:1988cp,Peebles:2002gy}, and the origin of the current
accelerated expansion of the Universe \cite{Riess:1998cb}. Therefore, there
are enough motivations for modifying general relativity. For instance, a
modification of general relativity can provide a plausible way to explain
the late-time acceleration of the Universe without a dark energy component
\cite{Clifton:2011jh, Carroll:2004de}.

In the context of modern particle physics, general relativity can be
considered as a unique theory of a massless Lorentz-invariant spin-2
particle (i.e., the graviton) in four dimensions \cite{Weinberg:1965rz}.
Actually, to find alternative theories to general relativity, we need to
break one of the underlying assumptions. One possible way is, breaking
Lorentz invariance where these theories contain additional degrees of
freedom \cite{Mattingly:2005re}. Another possible way is, maintaining
Lorentz invariance and considering gravity as a representation of a higher
spin \cite{Vasiliev:1995dn}. Here we consider a valuable alternative
theory, the massive gravity theory. In this theory, the gravity is
propagated by a spin-2 massive graviton.

The mass of graviton in massive gravity theory determines the speed of
gravitational wave propagation. Recently, gravitational waves have been
detected from merging of two neutron stars
\cite{TheLIGOScientific:2017qsa}. These events give us an opportunity to
have electromagnetic waves besides the gravitational waves. One
significance of these observations lays in the fact that the speed of these
waves can be compared and can give us the constraints on the modified
gravity theories.

It is well known that the massive gravity theory was introduced by Fierz
and Pauli in 1939 \cite{Fierz:1939ix}. They found the unique
Lorentz-invariant linear theory without ghost. In the following, the
massive gravity theory has undergone tremendous changes throughout decades.
The striking changes are discoveries of the van~Dam-Veltman-Zakharov (vDVZ)
discontinuity \cite{vanDam:1970vg,Zakharov:1970cc}, the Vainshtein
mechanism \cite{Vainshtein:1972sx}, and Boulware-Deser ghost
\cite{Boulware:1973my}. Eventually, the de~Rham-Gabadadze-Tolley (dRGT)
theory, which is a fully nonlinear massive gravity without Boulware-Deser
ghost, was introduced in 2010 by de Rham, Gabadadze and Tolley
\cite{deRham:2010ik,deRham:2010kj}.

It is expected that the extended massive gravity theories can explain the
cosmic acceleration without dark energy. As all homogeneous and isotropic
cosmological solutions in dRGT theory are unstable \cite{DeFelice:2012mx},
there are two alternative approaches. In the first approach, either
homogeneity or isotropy of background can be broken \cite{DAmico:2011eto,
Gumrukcuoglu:2012aa, DeFelice:2013awa}. In the second approach, we can
consider the extra degrees of freedom such as an extra scalar field or an
additional spin-2 field \cite{DAmico:2012hia, Huang:2012pe, Hassan:2011zd,
Hinterbichler:2012cn}. The quasi-dilaton massive gravity theory is
classified in the second approach. This theory introduces an extra scalar
degree of freedom to the dRGT theory \cite{DAmico:2012hia}. Meanwhile, it
should be pointed out that there are efforts to extend the quasi-dilaton
massive gravity theory \cite{Gumrukcuoglu:2013nza, Mukohyama:2014rca,
DeFelice:2013tsa}. In this paper, we propose a new extended quasi-dilaton
massive gravity theory which is achieved by adding a Gauss-Bonnet term.

Actually, Gauss-Bonnet theory was introduced by Lanczos
\cite{Lanczos:1938sf}, and Lovelock studied more details of this theory
\cite{Lovelock:1971yv}. It is interesting to note that the Gauss-Bonnet
gravity includes curvature-squared terms which have quadratic order of
derivatives with respect to the metric
\cite{Boulware:1985wk,Zumino:1985dp}. Generally, it can be mentioned that
this theory is ghost-free and can solve some problems in general relativity
\cite{Nojiri:2018ouv,Stelle:1977ry}. In addition, we point out that
Gauss-Bonnet theory arises from the low-energy limit of heterotic string
theories \cite{Zwiebach:1985uq,Gross:1986iv}. It is worth noting that there are valuable investigations which have something to do with the inflation in Gauss-Bonnet theory and cosmological perturbations \cite{Guo:2010jr,Koh:2014bka,Myung:2014jha,Kucukakca:2020ydp,Kawai:1998ab,Kawai:1999pw,Satoh:2008ck,Satoh:2007gn,Hikmawan:2015rze}. Actually in Refs. \cite{Guo:2010jr,Koh:2014bka}, the slow-roll inflation with a nonminimally coupled Gauss-Bonner term, was investigated analytically and numerically. They analyzed and constrained their models and results by the 7-year Wilkinson Microwave Anisotropy Probe, Planck and BICEP2 data, respectively.

There has been a trend toward cosmological and perturbation analysis of
extended quasi-dilaton massive gravity theories. For example, the
cosmological perturbations in extended massive gravity were studied in
Ref.~\cite{Gumrukcuoglu:2013nza}; the self-accelerated solutions in
quasi-dilaton massive gravity were purposed in
Ref.~\cite{Gabadadze:2014kaa}; the stability of self-accelerating solutions
in that theory in the presence of matter was investigated in
Ref.~\cite{Motohashi:2014una}; other investigations can be found in
Refs.~\cite{Mukohyama:2014rca, Gumrukcuoglu:2016hic, Mukohyama:2013raa,
Kahniashvili:2014wua,Gannouji:2013rwa}.

The goal of this paper is introducing a new extension of quasi-dilaton
massive gravity theory which is achieved by adding the Gauss-Bonnet term.
In this paper, we introduce the cosmological analysis and tensor
perturbation in order to calculate the mass of graviton. Actually, we
analyse the constraint on the mass of graviton according to this new
action. The paper is organized as follows. In Sec. \ref{sec:1} we introduce
the new action which contains the quasi-dilaton massive gravity and
Gauss-Bonnet terms. In following stage, we derive the background equations
of motion and self-accelerating solutions elaborately. In Sec. \ref{sec:2}
we perform perturbation analysis for determining the mass of graviton in
this theory and we discuss the graviton mass bounds in comparison with
gravitational-wave data. Finally, in Sec. \ref{sec:3} we conclude with a
discussion.

%---------------------------------------------------------------------
\section{\label{sec:1}Cosmological Background}
%---------------------------------------------------------------------

In this section, we review the quasi-dilaton dRGT massive gravity theory
which is extended by the Gauss-Bonnet term, and we discuss the evolution of
a cosmological background. The action includes Planck mass $\PlanckMass$,
the Ricci scalar $R$, the cosmological constant $\Lambda$, a dynamical
metric $g_{\mu\nu}$ and its determinant $\sqrt{-g}$. The action is given by
%--
\begin{eqnarray}\label{Action}
S=\frac{\PlanckMass^{2}}{2}\int d^{4} x \Bigg\{\sqrt{-g}\bigg[R-2\Lambda
+2{m}_{g}^{2}U(\mathcal K)\nonumber\\
-\frac{\omega}{\PlanckMass^{2}}g^{\mu\nu}\partial_{\mu}\sigma \partial_{\nu}
\sigma +\xi (\sigma) G(R)\bigg]\Bigg\}.
\end{eqnarray}
%--
In the following, we introduce two main parts---namely $U({\cal K})$ and
$G(R)$---of this action separately.

%---------------------------------------------------------------------
\subsection{Quasi-dilaton massive gravity term}\label{subsec:1}
%---------------------------------------------------------------------

In the first part, we start out with introducing the quasi-dilaton massive
gravity theory which includes the massive graviton term and the quasi-dilaton
term \cite{DAmico:2012hia}. Let us now introduce these two parts as a
single theory. It is clear that the mass of graviton comes up with the
potential $U$ which consists of three parts.
%--
\begin{equation}\label{Upotential1}
U(\mathcal{K})=U_{2}+\alpha_{3}U_{3}+\alpha_{4}U_{4},
\end{equation}
%--
where $\alpha_3$ and $\alpha_4$ are dimensionless free parameters of the
theory. $U_i$ ($i=2,3,4$) is given by,
%--
\begin{eqnarray}\label{Upotential2}
 U_{2}&=&\frac{1}{2}\big([\mathcal{K}]^{2}-[\mathcal{K}^{2}]\big),
 \nonumber\\
 U_{3}&=&\frac{1}{6}\big([\mathcal{K}]^{3}-3[\mathcal{K}][\mathcal{K}^{2}]+2[\mathcal{K}^{3}]\big),
 \nonumber\\
 U_{4}&=&\frac{1}{24}\big([\mathcal{K}]^{4}-6[\mathcal{K}]^{2}[\mathcal{K}^{2}]+8[\mathcal{K}][\mathcal{K}^{3}]+3[\mathcal{K}^{2}]^2\nonumber\\&&-6[\mathcal{K}^{4}]\big),
\end{eqnarray}
%--
where the quantity ``$[\cdot]$'' is interpreted as the trace of the tensor
inside brackets. It is essential to mention that the building block tensor
$\mathcal{K}$ is defined as
%--
\begin{equation}\label{K}
\mathcal{K}^{\mu}_{\nu} = \delta^{\mu}_{\nu} -
e^{\sigma/\PlanckMass}\sqrt{g^{\mu\alpha}f_{\alpha\nu}},
\end{equation}
%--
where $ f_{\alpha\nu}$ is the fiducial metric, which is defined through
%--
\begin{equation}\label{7}
f_{\alpha\nu}=\partial_{\alpha}\phi^{c}\partial_{\nu}\phi^{d}\eta_{cd}.
\end{equation}
%--
Here $g^{\mu\nu} $ is the physical metric, $\eta_{cd}$ is the Minkowski
metric with $c,d= 0,1,2,3$ and $\phi^{c}$ are the Stueckelberg fields which
are introduced to restore general covariance. Also, it is important to note
that $\sigma$ is the quasi-dilaton scalar and $\omega$ is a dimensionless
constant. Moreover, the theory is invariant under a global dilation
transformation, $\sigma\rightarrow\sigma+\sigma_{0}$.

According to our cosmological application purpose, we adopt the
Friedman-Lema\^itre-Robertson-Walker (FLRW) Universe. So, the general
expression of the corresponding dynamical and fiducial metrics are given as
follows,
%--
\begin{align}
\label{DMetric}
g_{\mu\nu}&={\rm diag} \left[-N^{2},a^2,a^2,a^2 \right], \\
\label{FMetric} 
f_{\mu\nu}&={\rm diag} \left[-\dot{f}(t)^{2},1,1,1 \right].
\end{align}
%--
Here it is worth pointing out that $N$ is the lapse function of the
dynamical metric, and it is similar to a gauge function. Also, it is clear
that the scale factor is represented by $a$, and $\dot{a}$ is the
derivative with respect to time. Furthermore, the lapse function relates
the coordinate-time $dt$ and the proper-time $d\tau$ via $d\tau=Ndt$
\cite{Scheel:1994yr,Christodoulakis:2013xha}. Function $f(t)$ is the
Stueckelberg scalar function whereas $\phi^{0}=f(t)$ and
$\frac{\partial\phi^{0}}{\partial t}=\dot{f}(t)$ \cite{ArkaniHamed:2002sp}.

Therefore, the Lagrangian of the quasi-dilaton massive gravity in FLRW
cosmology is
%--
\begin{eqnarray}
\mathcal{L}_{\rm QD}=&&\PlanckMass^{2}\bigg[-3\frac{a\dot{a}^{2}}{N}-\Lambda{a}^{3}N\bigg]+m_{g}^{2}\PlanckMass^{2}\Bigg\lbrace Na^{3}(X-1)\nonumber\\
&&\times\bigg[3(X-2)-(X-4)(X-1)\alpha_{3}-(X-1)^{2}\alpha_{4}\bigg]\nonumber\\
&&+\dot{f}(t)a^{4}X(X-1)\bigg[3-3(X-1)\alpha_{3}+(X-1)^{2}\alpha_{4}\bigg]\Bigg\rbrace\nonumber\\
&&+\frac{\omega a^{3}}{2N}\dot{\sigma}^{2},
\end{eqnarray}
%--
where
%--
\begin{equation}\label{XX}
    X\equiv\frac{e^{\sigma/\PlanckMass}}{a}.
\end{equation}
%--

%---------------------------------------------------------------------
\subsection{Gauss-Bonnet term}\label{subsec2}
%---------------------------------------------------------------------

Here, we introduce the Gauss-Bonnet term which we add to the quasi-dilaton
massive gravity theory. This term consists of the Gauss-Bonnet invariant,
$G(R)=R_{\mu\nu\gamma\delta}R^{\mu\nu\gamma\delta}-4R_{\mu\nu}R^{\mu\nu}+R^2$,
and a coupling function $\xi(\sigma)$. It should be noted that in $D=4$ if we
consider $\xi$ as a dimensionless coupling constant instead of a coupling
function $\xi(\sigma)$, the Gauss-Bonnet term does not contribute to the
gravitational dynamics. The reason of this lays in the fact that the
Gauss-Bonnet invariant is a total derivative \cite{Glavan:2019inb}. So, in
this paper, we adopt the coupling function $\xi(\sigma)$ similar to
Ref.~\cite{Chatzarakis:2019fbn}. Using integration by parts, we can convert
the second derivative terms into the first order derivatives. The part of
the Lagrangian which is related to the Gauss-Bonnet term is
%--
\begin{eqnarray}\label{LGB}
\mathcal{L}_{\rm GB}=\frac{\PlanckMass^{2}}{2}\sqrt{-g}\xi(\sigma)G(R)\rightarrow-4 \PlanckMass^{2}\frac{\dot{a}^{3}}{{N}^{3}}\xi'(\sigma)\dot{\sigma},
\end{eqnarray}
%--
where the last expression shows its form in the FLRW background.
As a result, the total Lagrangian includes two parts,
%--
\begin{eqnarray}\label{TotalL}
\mathcal{L}=\mathcal{L}_{\rm QD}+\mathcal{L}_{\rm GB}.
\end{eqnarray}
%--
So, the point-like Lagrangian for cosmology is
%--
\begin{eqnarray}
\mathcal{L}&=&\PlanckMass^{2}\bigg[-3\frac{a\dot{a}^{2}}{N}-\Lambda{a}^{3}N\bigg]
+m_{g}^{2}\PlanckMass^{2}\bigg[N a^{3}(X-1)\nonumber\\
&&\times\big[3(X-2)-(X-4)(X-1)\alpha_{3}-(X-1)^{2}\alpha_{4}\big]\nonumber\\
&&+\dot{f}(t)a^{4}X(X-1)\big[3-3(X-1)\alpha_{3}+(X-1)^{2}\alpha_{4}\big]\bigg]\nonumber\\
&&+\frac{\omega a^{3}}{2N}\dot{\sigma}^{2}-4\PlanckMass^{2}\frac{\dot{a}^{3}}{N^{3}}\xi'(\sigma)\dot{\sigma}.
\end{eqnarray}
%--
In order to simplify expressions later, we define
%--
\begin{equation}
H\equiv\frac{\dot{a}}{Na}.
\end{equation}
%--

%---------------------------------------------------------------------
\subsection{Background equations of motion}\label{subsec4}
%---------------------------------------------------------------------

In order to achieve a constraint equation we should take the unitary gauge
into consideration, which means that we choose $f(t)=t$. The significance of
the unitary gauge lays in the fact that on the classical level the
unphysical fields could be eliminated from the Lagrangian with use of gauge
transformations \cite{GrosseKnetter:1992nn}. In this procedure, a
constraint equation can be derived by varying with respect to $f$. So, that
equation is given by
%--
\begin{eqnarray}\label{Cons}
\hspace{-0.5cm} m_{g}^{2} \PlanckMass^{2}\frac{d}{dt}\bigg[ && a^{4}X(X-1)\nonumber\\
\hspace{-0.5cm} && \times[3-3(X-1)\alpha_{3}+(X-1)^{2}\alpha_{4}]\bigg]=0.
\end{eqnarray}
%--
In this stage, the Friedman equation is achieved by varying with respect to
the lapse $N$,
%--
\begin{eqnarray}\label{EqN}
    3H^{2}-\Lambda && -\frac{\omega}{2}\Big(H+\frac{\dot{X}}{XN}\Big)^{2} -m_{g}^{2}(X-1)\bigg[-3(X-2)\nonumber\\
    && +(X-4)(X-1)\alpha_{3}+(X-1)^{2}\alpha_{4}\bigg]\nonumber\\
    && +12H^{3}\PlanckMass\Big(H +\frac{\dot{X}}{XN}\Big)\xi'(\sigma)=0.
\end{eqnarray}
%--
The equation of motion for $\sigma$ is
%--
\begin{eqnarray}\label{EqSig}
12\PlanckMass H^{2} && \Big(H^{2}+\frac{\dot{H}}{N}\Big)\xi'(\sigma)-\frac{\omega}{\PlanckMass}\bigg[3H\frac{\dot{\sigma}}{N}
+\frac{1}{N}\frac{d}{dt}\Big(\frac{\dot{\sigma}}{N}\Big)\bigg]
\nonumber\\
&&+3m_{g}^{2}X\bigg\lbrace\big(2X-3+r(2X-1)\big) \nonumber\\
&&+(X-1)\bigg[\alpha_{3}\big(3-X+r(1-3X)\big)\nonumber\\
&&-\frac{1}{3}\alpha_{4}(X-1)\big(3+r(1-4X)\big)\bigg]\bigg\rbrace =0,
\end{eqnarray}
%--
where
%--
\begin{eqnarray}
r\equiv\frac{a}{N}.
\end{eqnarray}
%--
Using the notation in Eq. (\ref{XX}), the following equations can be
derived
%--
\begin{equation}
\frac{\dot{\sigma}}{N\PlanckMass}= H+\frac{\dot{X}}{NX}, \qquad \frac{\ddot{\sigma}}{\PlanckMass}=\frac{d}{dt}\Big(NH+\frac{\dot{X}}{X}\Big),
\end{equation}
%--
and the last equation of motion could be obtained by varying with respect
to the scale factor $a$,
%--
\begin{eqnarray}\label{Eqa}
3H^{2} && -\Lambda +\frac{\omega}{2 \PlanckMass^{2}}\Big(\frac{\dot{\sigma}}{N}\Big)^{2}+4\frac{d}{dt}\Big(H^{2}\xi'\frac{\ddot{\sigma}}{N^{2}}\Big)\nonumber\\
&&-4\frac{H^{2}}{N}\xi'\bigg[\frac{d}{dt}\Big(\frac{\dot{\sigma}}{N}\Big)-\Big(\frac{\ddot{\sigma}}{N}\Big)-2H\dot{\sigma}\bigg]+2\frac{\dot{H}}{N}\nonumber\\
&&+m_{g}^{2}\bigg\lbrace 1+rX(2X-3)+(X-1)\bigg[X -5 \nonumber\\
&& -\alpha_{3}\big(4-2X+rX(X-3)\big)\nonumber\\
&&-\alpha_{4}(X-1)(rX-1)\bigg]\bigg\rbrace =0.
\end{eqnarray}
%--
In the last part of this subsection, it should be noted that the Stuckelberg field $f$ introduces time reparametrization invariance. So, there is a Bianchi identity which relates the four equations of motion,
%--
\begin{eqnarray}
\frac{\delta S}{\delta \sigma}\dot{\sigma}+\frac{\delta S}{\delta f}\dot{f}-N\frac{d}{dt}\frac{\delta S}{\delta N}+\dot{a}\frac{\delta S}{\delta a}=0.
\end{eqnarray}
%--
So, one equation is redundant and can be eliminated.

%---------------------------------------------------------------------
\subsection{Self-accelerating background solutions}\label{subsec5}
%---------------------------------------------------------------------

In this step, we want to discuss solutions. It could be started with
the Stueckelberg constraint in Eq. (\ref{Cons}). After integrating the
equation we have
%--
\begin{eqnarray}\label{Self}
X(X-1)\bigg[3-3(X-1)\alpha_{3}+(X-1)^{2}\alpha_{4}\bigg] \propto a^{-4} .
\end{eqnarray}
%--
It would be suitable to mention that the constant solutions of $X$ lead to
the effective energy density and behave similar to a cosmological constant.
If we consider an expanding universe, according to the $a^{-4}$ behavior
in Eq.~(\ref{Self}), the right-hand side of that equation will decrease.
Therefore, after a long enough time, $X$ leads to a constant value, $X_{\rm
SA}$, which is a root of the left-hand side of Eq.~(\ref{Self}).

One of the solutions for Eq. (\ref{Self}) is $X=0$ which leads to
$\sigma\rightarrow-\infty$. Meanwhile, this solution multiplies to the
perturbations of the auxiliary scalars which means that we encounter strong
coupling in the vector and scalar sectors. Thus, in order to avoid strong
coupling, we discard this solution \cite{DAmico:2012hia}. So, we are left with,
%--
\begin{equation}
\hspace{-0.4cm}
(X-1)\big[3-3(X-1)\alpha_{3}+(X-1)^{2}\alpha_{4}\big]\bigg|_{X=X_{\rm SA}}=0.
\end{equation}
%--
An obvious solution is $X=1$ which leads to a vanishing cosmological constant
and because of inconsistency it is unacceptable. So, this solution should
be discarded too \cite{DAmico:2012hia}.

As a result, the two remaining solutions of Eq. (\ref{Self}) are
%--
\begin{equation}\label{XSA}
X_{\rm SA}^{\pm}=\frac{3\alpha_{3}+2\alpha_{4}\pm\sqrt{9\alpha_{3}^{2}-12\alpha_{4}}}{2\alpha_{4}}.
\end{equation}
%--
The Friedman equation (\ref{EqN}) could be written in a different form,
%--
\begin{eqnarray}\label{EqFr}
\bigg(3-\frac{\omega}{2}+12 \PlanckMass \xi'(\sigma)H^{2}\bigg){H^{2}}=\Lambda + \Lambda_{\rm SA}^{\pm}.
\end{eqnarray}
%--
Considering self-accelerating solutions, in the case of
$\xi'(\sigma)=0$, a condition on the parameter $\omega$ is provided by the
Friedman equation (\ref{EqFr}). So, we need to consider $\omega<6$ to keep
the left hand side of Eq.~(\ref{EqFr}) positive. The importance of this
issue lays in the fact that when we add ordinary matters to the right-hand
side, throughout the matter dominated era, we will have the standard
cosmology.

It is worth mentioning that the effective cosmological constant from the
mass term is
%--
\begin{eqnarray}
\Lambda_{\rm SA}^{\pm}\equiv m_{g}^{2}(X_{\rm SA}^{\pm}-1)\bigg[ && -3X_{\rm SA}^{\pm} +6+(X_{\rm SA}^{\pm}-4)(X_{\rm SA}^{\pm}-1)\alpha_{3}\nonumber\\
&&+(X_{\rm SA}^{\pm}-1)^{2}\alpha_{4}\bigg].
\end{eqnarray}
%--
According to Eq. (\ref{XSA}), the above equation can be written as 
%--
\begin{eqnarray}
\Lambda_{\rm SA}^{\pm}=\frac{3m^{2}_{g}}{2\alpha^{3}_{4}}\bigg[9\alpha^{4}_{3}\pm 3\alpha^{3}_{3}\sqrt{9\alpha^{2}_{3}-12\alpha_{4}}-18\alpha^{2}_{3}\alpha_{4}\nonumber\\\mp 4\alpha_{3}\alpha_{4}\sqrt{9\alpha^{2}_{3}-12\alpha_{4}}+6\alpha^{2}_{4}\bigg].
\end{eqnarray}
%--
Therefore, $H^{2}$ is obtained via Eq. (\ref{EqFr}),
%--
\begin{eqnarray}\label{H2}
H^{2}=\frac{1}{24 \PlanckMass \xi'(\sigma)} \Bigg\{ && -(3-\frac{\omega}{2})\nonumber\\
&& \hspace{-1.8cm} \mp \bigg[(3-\frac{\omega}{2})^{2} +48 \PlanckMass (\Lambda +\Lambda_{\rm SA}^{\pm})\xi'(\sigma)\bigg]^{\frac{1}{2}}\Bigg\}.
\end{eqnarray}
%--
Therefore, for the self-accelerating solutions, there are two cases. 

In the first case, $\xi'(\sigma)$ is a constant so
$\xi^{\prime\prime}(\sigma)$ is equal to zero. Therefore, from
Eq.~(\ref{EqSig}) we have,
%--
\begin{eqnarray}\label{rS1}
r_{\rm SA1}=1+\frac{H^{2}\big[\omega -4 \PlanckMass H^{2}\xi'(\sigma)\big]}{m_{g}^{2}X_{\rm SA}^{2\pm}\big(-2-\alpha_{3}+\alpha_{3}X_{\rm SA}^{\pm}\big)}.
\end{eqnarray}
%--
It is important to note that, in this case, we can consider
$\xi'(\sigma)=\xi_{0}$ where $\xi_{0}$ is a constant parameter. In the
following, $\bar{\Lambda}$ is redefined as $\bar{\Lambda}=48 \PlanckMass
(\Lambda +\Lambda_{\rm SA}^{\pm})\xi_{0}$. Therefore, in order to keep the
right-hand side of Eq.~(\ref{H2}) positive, the below conditions should be
satisfied in ``$\pm$'' cases respectively:
%--
\begin{itemize}
    \item In the case of ``$-$'' sign, we should have $\bar{\Lambda}<0$ and
    $\omega\geq 6+2\sqrt{-\bar{\Lambda}}$; in other words, $\xi_{0}$ has to be
    negative.
    \item In the case of ``$+$'' sign, there are two conditions: (a) it can be
    considered $\bar{\Lambda}<0$, $\omega\geq 6+\sqrt{-\bar{\Lambda}}$, and also
    $\xi_{0}$ should be smaller than zero; (b) if we consider
    $\bar{\Lambda}\geq 0$, the right-hand side of Eq.~(\ref{H2}) is positive
    for any $\omega$.
\end{itemize}
%--

In the second case, $\xi'(\sigma)$ is an arbitrary function so
$\xi^{\prime\prime}(\sigma)$ is not zero. In the following, we calculated
$\dot{H}$ using Eq. (\ref{EqFr}) and we substitute it into Eq. (\ref{rS1}).
As a result, we obtain,
%--
\begin{eqnarray}\label{rSA2}
&& \hspace{-0.5cm} r_{\rm SA2}=1+\frac{H^{2}}{m_{g}^{2}X_{\rm SA}^{2\pm}(-2-\alpha_{3}+\alpha_{3}X_{\rm SA}^{\pm})}\nonumber\\ 
&& \hspace{-0.35cm} \times \bigg[\omega -4 \PlanckMass \xi{'}(\sigma)\bigg(H^{2}+\frac{12H^{4}\PlanckMass^2\xi{''}(\sigma)}{-6+\omega -48H^{2}\PlanckMass\xi{'}(\sigma)}\bigg)\bigg].
\end{eqnarray}
%--
Actually, we have used the Stuckelberg equation (\ref{Self}) in order to
eliminate $\alpha_{4}$. Finally, we should take this into account, and if
we consider $\xi (\sigma)=0$, $r_{\rm SA1}$ and $r_{\rm SA2}$ convert to
the equation in Ref.~\cite{Gumrukcuoglu:2013nza} in its unexpanded form.

%---------------------------------------------------------------------
\section{Tensor perturbation}\label{sec:2}
%---------------------------------------------------------------------

In this section, we would like to analyse tensor perturbation in order to
calculate the mass of graviton for our theory which we introduced in
the previous section.

In order to find the action for quadratic perturbation, the physical metric
is expanded in small fluctuation, $\delta g_{\mu\nu}$, around a solution
$g_{\mu\nu}^{(0)}$,
%--
\begin{equation}
g_{\mu\nu}=g_{\mu\nu}^{(0)}+\delta g_{\mu\nu}.
\end{equation}
%--
In the following analysis, we keep terms to quadratic order in $\delta
g_{\mu\nu}$. As we demonstrate all analysis in the unitary gauge, there are
not any problems concerning the form of gauge invariant combinations.
Moreover, we write the actions expanded in the Fourier domain with plane
waves, i.e., $\vec{\nabla}^{2}\rightarrow -k^{2}$, $d^{3}x\rightarrow
d^{3}k$. We raise and lower the spatial indices on perturbations by
$\delta^{ij}$ and $\delta_{ij}$. It should be mentioned that we would like
to consider $N=1$ which means that the derivatives are with respect to
time.

We start by considering tensor perturbations around the background,
%--
\begin{equation}
\delta g_{ij}=a^{2}h_{ij}^{\rm TT},
\end{equation}
%--
where
%--
\begin{equation}
    \partial^{i}h_{ij}=0 \quad {\rm and} \quad g^{ij}h_{ij}=0.
\end{equation}
%--
Therefore, the action quadratic in $h_{ij}$ is 
%--
\begin{eqnarray}
S&=&\frac{\PlanckMass^{2}}{8}\int d^{3}k \, dt \, a^{3}\Bigg[ \bigg(1-4H\xi{'}(\sigma)\bigg)\dot{h}^{ij}\dot{h}_{ij}\nonumber\\
&&-\bigg(\frac{k^{2}}{a^{2}}\big[1-4\xi{''}(\sigma)\big]+M_{\rm GW}^{2}\bigg)h^{ij}h_{ij}\Bigg].
\end{eqnarray}
%--
As we have $r_{\rm SA1}$ and $r_{\rm SA2}$, in order to calculate the
dispersion relation of gravitational waves we have two cases. In the first
case, we obtain $\alpha_{3}$ using Eq.~(\ref{rS1}) and $\alpha_{4}$ using
Eq.~(\ref{XSA}). So, in this case, the dispersion relation of gravitational
waves is
%--
\begin{eqnarray}
M^{2}_{\rm GW_{1}}&=&4\dot{H}+6H^{2}-2\Lambda -16\xi_{0}H\big(\dot{H}+H^{2}\big) \nonumber\\
&&+\omega H^{2}+\Upsilon_{1}, \label{eq:M2:GW1}
\end{eqnarray}
%--
where
%--
\begin{widetext}
\begin{eqnarray}
\Upsilon_{1}&=&\frac{1}{(r_{\rm SA1}-1)(X_{\rm SA}^{\pm}-1)(X_{\rm
SA}^{\pm})^{2}} \Bigg\lbrace\omega H^{2}\bigg[X_{\rm SA}^{\pm}(X_{\rm SA}^{\pm}
-3)(r_{\rm SA1}X_{\rm SA}^{\pm}-2)-2\bigg] \nonumber\\
&& \hspace{-1cm} +m_{g}^{2}(r_{\rm SA1}-1)X_{\rm SA}^{\pm}\bigg[X_{\rm
SA}^{\pm}\bigg(6+X_{\rm SA}^{\pm}[X_{\rm SA}^{\pm}(1+r_{\rm
SA1})-6]\bigg)-2\bigg]
+4H^{4}{\PlanckMass}\xi_{0}\bigg[2-X_{\rm SA}^{\pm}(X_{\rm
SA}^{\pm}-3)(r_{\rm SA1}X_{\rm SA}^{\pm}-2)\bigg]\Bigg\rbrace.
\end{eqnarray}
\end{widetext}
%--
In the second case, $\alpha_{3}$ can be gotten from Eq.
(\ref{rSA2}), and similar to the last case we obtain $\alpha_{4}$ from
Eq.~(\ref{XSA}). Therefore, the dispersion relation of gravitational waves is
obtained for the second case,
%--
\begin{eqnarray}
M^{2}_{\rm GW_{2}}&=&4\dot{H}+6H^{2}-2\Lambda
+8\xi{''}(\sigma)H^{2}\nonumber\\
&&-16\xi{'}(\sigma)H\big(\dot{H}+H^{2}\big)+\omega H^{2}+\Upsilon_{2},
\label{eq:M2:GW2}
\end{eqnarray}
%--
where
%--
% \begingroup\makeatletter\def\f@size{8}\check@mathfonts
\begin{widetext}
\begin{eqnarray}
\Upsilon_{2}&=&\frac{1}{(r_{\rm SA2}-1)(X_{\rm SA}^{\pm}-1)(X_{\rm
SA}^{\pm})^{2}} \, \frac{1}{\omega -48H^{2}\PlanckMass\xi{'}(\sigma) -6}
\Bigg\lbrace m_{g}^{2}(\omega -6)(r_{\rm SA2}-1)X_{\rm
SA}^{\pm}\bigg[X_{\rm SA}^{\pm}\big[6+X_{\rm SA}^{\pm}(X_{\rm
SA}^{\pm}(r_{\rm SA2}+1)-6)\big]-2\bigg]\nonumber\\
&&-4(13 \omega -6)H^{4}\PlanckMass\xi{'}(\sigma)\bigg[X_{\rm
SA}^{\pm}\big(X_{\rm SA}^{\pm}-3\big)\big(r_{\rm SA2}X_{\rm SA}^{\pm}
-2\big)-2\bigg]+H^{2}\bigg[\omega\big(\omega -6\big)\bigg(X_{\rm
SA}^{\pm}\big(X_{\rm SA}^{\pm}-3\big)\big(r_{\rm SA2}X_{\rm SA}^{\pm}
-2\big)-2\bigg) \nonumber\\
&&-48m_{g}^{2}\PlanckMass\xi{'}(\sigma)(X_{\rm SA}^{\pm})^{2}(r_{\rm
SA2}-1)\bigg(X_{\rm SA}^{\pm} \big[X_{\rm SA}^{\pm}(r_{\rm
SA2}+1)-6\big]-2\bigg)\bigg]\nonumber\\
&&-48H^{6}\PlanckMass^{2}\xi{'}(\sigma)\bigg(\PlanckMass\xi{''}(\sigma)-4\xi{'}(\sigma)\bigg)\bigg[X_{\rm
SA}^{\pm}\bigg(r_{\rm SA2}X_{\rm SA}^{\pm}-2\bigg)\big(X_{\rm
SA}^{\pm}-3\big)-2\bigg]\Bigg\rbrace.
\end{eqnarray}
\end{widetext}
% \endgroup
%--
As we mentioned before, we eliminate $\alpha_{3}$ and $\alpha_{4}$ using
Eq.~(\ref{XSA}), and Eqs.~(\ref{rS1}--\ref{rSA2}). It can be pointed out
that if the mass square of gravitational waves is positive, the stability
of long-wavelength gravitational waves is guaranteed. On the other hand, if
it is negative, it should be tachyonic. Therefore, as the mass of tachyon
is of the order of Hubble scale, the instability should take the age of the
Universe to develop.

The main results of this section are the modified dispersion relations of
gravitational waves, given in Eq.~(\ref{eq:M2:GW1}) and
Eq.~(\ref{eq:M2:GW2}). They represent the propagation of gravitational
perturbations in the FLRW cosmology in the Gauss-Bonnet quasi-dilaton
massive gravity. In principle, the propagation can be tested with
cosmological events, notably by gravitational wave observations. These
modifications will introduce extra contribution to the phase evolution of
gravitational waveform~\cite{Will:1997bb, Mirshekari:2011yq}, and to be
detected with the accurate matched-filtering techniques in the data
analysis.

After the first discovery of gravitational waves in a merging binary black
hole (the so-called GW150914) by the LIGO/Virgo Collaboration, tests of
graviton mass are revived~\cite{TheLIGOScientific:2016src,
LIGOScientific:2019fpa, Abbott:2020jks, Shao:2020shv}. The latest
constraint on the graviton mass from the combination of gravitational wave
events from the first and second gravitational wave transient catalogs is
$m_{g} \leq 1.76 \times 10^{-23} \, \mathrm{eV} / c^{2}$ at 90\%
credibility~\cite{Abbott:2020jks}. The corresponding Compton wavelength is
still much smaller than the Hubble scale, thus the relevance to modified
cosmology is restricted at present. Nevertheless, with future prospects in
mind, we shall keep testing this important aspect of gravitation with more
and more gravitational events at different wavelengths, notably with future
space-based gravitational-wave detectors which are more sensitive to the
graviton mass~\cite{Will:1997bb}. 

In particular, modified propagation of gravitational waves in the
cosmological setting was investigated by Nishizawa and Arai in a
parametrized framework, considering a running Planck mass, a modified
speed for gravitation, as well as anisotropic source
terms~\cite{Nishizawa:2017nef, Arai:2017hxj, Nishizawa:2019rra}. Our
results here, representing theory-specific analysis in Gauss-Bonnet
quasi-dilaton massive gravity, are complementary to their work, and will
provide a target for future detailed analysis.

%---------------------------------------------------------------------
\section{Conclusion}\label{sec:3}
%---------------------------------------------------------------------
In this work, we have presented a new extension of quasi-dilaton massive gravity theory which is constructed by adding the Gauss-Bonnet term. As the quasi-dilaton massive gravity and its extensions have a rich phenomenology, we have been motivated to investigate some cosmological analysis of Gauss-Bonnet quasi-dilaton massive gravity.

At the first, we have introduced the details of the new action and total Lagrangian. We also presented the full set of equations of motion for a FLRW background. Notice that the investigation of extended massive gravity is important in order to understand the late-time acceleration of the Universe. Therefore, we have discussed the self-accelerating background solutions elaborately. We have provided a way to explain the late-time acceleration of the Universe within the Gauss-Bonnet quasi-dilaton massive gravity.

To study the mass of graviton for the Gauss-Bonnet quasi-dilaton massive gravity theory, we have presented the tensor perturbation calculation and have shown the dispersion relation of gravitational waves for two cases. In other words, we have represented the propagation of gravitational perturbation in the FLRW cosmology in the Gauss-Bonnet quasi-dilaton massive gravity. Such an analysis will be a useful addition to probe alternative gravity theories in the era of gravitational waves. In addition, a detailed direct comparison with observational data (e.g., from type Ia supernovae and the cosmic microwave background) for the late-time acceleration of the Universe will be extremely interesting to check for the valid parameter space of the Gauss-Bonnet quasi-dilaton massive gravity theory in this work. It will also be useful for an insightful comparison with the canonical $\Lambda$CDM cosmology model. However, such a statistical study dedicated to data analysis is beyond the scope of the current paper, thus we leave it for future study.

At the end, we think that other possible extensions of quasi-dilaton massive gravity theory can be considered for future investigations.  
\\
%---------------------------------------------------------------------
\section*{Acknowledgements}
We are grateful to the anonymous referee for helpful comments.
This work has been supported by University of Tabriz, International and
Academic Cooperation Directorate, in the framework of TabrizU-300 program.
Also, A.R.K and S.K would like to thank A. Emir Gumrukcuolu and Nishant
Agarwal for useful comments. 
L.S was supported by the National Natural Science Foundation of China
(Grants No. 11975027, No. 11991053, No. 11721303), the Young Elite Scientists Sponsorship
Program by the China Association for Science and Technology (No. 2018QNRC001),
the National SKA Program of China (No. 2020SKA0120300), and the Max Planck
Partner Group Program funded by the Max Planck Society.
%---------------------------------------------------------------------

%\nocite{*}

%---------------------------------------------------------------------
\bibliography{apssamp}% Produces the bibliography via BibTeX.

%---------------------------------------------------------------------

\end{document}